\documentclass[twocolumn,showpacs,aps,prl,superscriptaddress,nofootinbib]{revtex4}

\usepackage{graphicx}
\usepackage{dcolumn}
\usepackage{amsmath}
\usepackage{epsfig}

\usepackage{maybemath}

\usepackage[italic]{hepparticles}

\usepackage{heppennames}

\def\BR                 {{\ensuremath{\cal B}\xspace}}

\newcommand{\forex}     {\mbox{\textsl{e.g.}}\xspace}

\def\xs         {\ensuremath{X_{s}}\xspace}    

\def\pizeta     {\Pgpz{}(\Pgh)\xspace}
\def\BB         {\ensuremath{\PB{}\PaB}\xspace}
\def\BzBzb      {\ensuremath{\PBz\PaBz}\xspace}
\def\BpBm       {\ensuremath{\PBp\PBm}\xspace}
\def\epem       {\ensuremath{\Pep\Pem}\xspace}

\def\mmg        {\ensuremath{\Pgm\Pgm\Pgg}\xspace}

\mathchardef\Upsilon="7107
\def\Y#1S{\ensuremath{\Upsilon{(#1S)}}\xspace}

\def\FourS {\Y4S}

 \def\eg         {\ensuremath{E_{\gamma} }\xspace}

 \def\egcms      {\ensuremath{E^{*}_{\gamma}}\xspace}

 \def\egb        {\ensuremath{E^{B}_{\gamma}}\xspace}

 \def\mxs        {\ensuremath{m_{X_{s}} }\xspace}

 \def\mupsq      {\ensuremath{\mu_{\pi}^2}\xspace}

 \def\mb         {\ensuremath{m_{b} }\xspace}

 \def\sig        {\ensuremath{\mathrm{S}}\xspace}

 \def\sigeff     {\ensuremath{\epsilon_\mathrm{sig}}\xspace}

 \def\bsg        {\ensuremath{\Pqb\to\Pqs\Pgg}}
 \def\bdg        {\ensuremath{\Pqb\to\Pqd\Pgg}\xspace}

 \def\bxsg       {\ensuremath{\PB \to X_{s} \Pgg}\xspace}
 \def\bxdg       {\ensuremath{\PB \to X_{d} \Pgg}\xspace}
 \def\bxsdg       {\ensuremath{\PB \to X_{s+d} \Pgg}\xspace}
 
 \def\bxclnu      {\ensuremath{\PB \to X_{c} \ell \nu }\xspace}

\def\efmbrest   {\ensuremath{\langle \eg \rangle}\xspace}

\def\varbrest   {\ensuremath{\langle (\eg - \langle \eg \rangle)^2 \rangle}\xspace}

\def\aveDelta#1 {\ensuremath{\langle \Delta_{total}#1 \rangle}\xspace}

\def\valerr#1#2#3 {\ensuremath{{#1}^{+#2}_{-#3}}\xspace}

\usepackage{relsize}
\usepackage{hyperref}
\def\babar{\mbox{\slshape B\kern-0.1em{\smaller A}\kern-0.1em
    B\kern-0.1em{\smaller A\kern-0.2em R}}}
\def\superb {\texorpdfstring{{Super\kern0.04em\emph{B}}\xspace}{SuperB\xspace}}

\newcommand{\tev}{\ensuremath{\mathrm{\,Te\kern -0.1em V}}\xspace}
\newcommand{\gev}{\ensuremath{\mathrm{\,Ge\kern -0.1em V}}\xspace}
\newcommand{\mev}{\ensuremath{\mathrm{\,Me\kern -0.1em V}}\xspace}
\newcommand{\kev}{\ensuremath{\mathrm{\,ke\kern -0.1em V}}\xspace}
\newcommand{\ev}{\ensuremath{\mathrm{\,e\kern -0.1em V}}\xspace}
\newcommand{\gevc}{\ensuremath{{\mathrm{\,Ge\kern -0.1em V\!/}c}}\xspace}
\newcommand{\mevc}{\ensuremath{{\mathrm{\,Me\kern -0.1em V\!/}c}}\xspace}
\newcommand{\gevcc}{\ensuremath{{\mathrm{\,Ge\kern -0.1em V\!/}c^2}}\xspace}
\newcommand{\mevcc}{\ensuremath{{\mathrm{\,Me\kern -0.1em V\!/}c^2}}\xspace}

\def\cm   {\ensuremath{{\rm \,cm}}\xspace}

\def\invfb   {\ensuremath{\mbox{\,fb}^{-1}}\xspace}

\def\mus  {\ensuremath{\rm \,\mus}\xspace}

\def\mus        {\ensuremath{\,\mu{\rm s}}\xspace}    

\def\to                 {\ensuremath{\rightarrow}\xspace}

\def\pep2{PEP-II}

\def\gsim{{~\raise.15em\hbox{$>$}\kern-.85em
          \lower.35em\hbox{$\sim$}~}\xspace}
\def\lsim{{~\raise.15em\hbox{$<$}\kern-.85em
          \lower.35em\hbox{$\sim$}~}\xspace}

\def\CP                {\ensuremath{C\!P}\xspace}

\def\acp        {\ensuremath{A_{\CP}}\xspace}

\def\amcp       {\ensuremath{A^{\mathrm{meas}}_{\CP}}\xspace}

\def\Vtd  {\ensuremath{|V_{td}|}\xspace}

\def\Vts  {\ensuremath{|V_{ts}|}\xspace}
\def\Vub  {\ensuremath{|V_{ub}|}\xspace}
\def\Vcb  {\ensuremath{|V_{cb}|}\xspace}

\def\bTosGamma{\HepProcess{\Pqb\to\Pqs\Pgg}\xspace}
\def\bTosplusdGamma{\HepProcess{\Pqb\to(\Pqs+\Pqd)\Pgg}\xspace}

\def\bbarTosbarplusdbarGamma{\HepProcess{\Paqb\to(\Paqs+\Paqd)\Pgg}\xspace}
\def\bTodGamma{\HepProcess{\Pqb\to\Pqd\Pgg}\xspace}

\def\bsg{\bTosGamma}

\def\bdg{\bTodGamma}

\def\lumibbpm   {\ensuremath {(382.8 \pm 4.2) \times 10^6}\xspace}

\def\onlumi     {\ensuremath { 347.1 \invfb }\xspace}

\def\offlumi    {\ensuremath {36.4 \invfb }\xspace}
\def\acpresult  {\ensuremath {0.057 \pm 0.063}\xspace}    
    
\def\PBFresultoneeightone {\ensuremath {(3.21 \pm 0.33)\times 10^{-4}}\xspace}

\newcommand{\BABARPubYear}    {12}
\newcommand{\BABARPubNumber}  {016}

\newcommand{\SLACPubNumber} {15180}
\newcommand{\LANLNumber} {000000}

\def\figurebox#1#2#3{%
    \def\arg{#3}%
    \ifx\arg\empty
    {\hfill\vbox{\hsize#2\hrule\hbox to #2{\vrule\hfill\vbox to #1{\hsize#2\vfill}\vrule}\hrule}\hfill}%
    \else
    {\hfill\epsfbox{#3}\hfill}%
    \fi}

\begin{document}

\preprint{\babar-PUB-\BABARPubYear/\BABARPubNumber} 
\preprint{SLAC-PUB-\SLACPubNumber} 
\preprint{hep-ex/\LANLNumber}

\begin{flushleft}

\babar-PUB-\BABARPubYear/\BABARPubNumber\\
SLAC-PUB-\SLACPubNumber\\
\end{flushleft}

\title{\boldmath \large Precision Measurement of the \bxsg Photon Energy Spectrum, \\
Branching Fraction, and Direct \CP Asymmetry  $\acp(\PB \to X_{s+d}\gamma$)
}

%
\author{J.~P.~Lees}
\author{V.~Poireau}
\author{V.~Tisserand}
\affiliation{Laboratoire d'Annecy-le-Vieux de Physique des Particules (LAPP), Universit\'e de Savoie, CNRS/IN2P3,  F-74941 Annecy-Le-Vieux, France}
\author{J.~Garra~Tico}
\author{E.~Grauges}
\affiliation{Universitat de Barcelona, Facultat de Fisica, Departament ECM, E-08028 Barcelona, Spain }
\author{A.~Palano$^{ab}$ }
\affiliation{INFN Sezione di Bari$^{a}$; Dipartimento di Fisica, Universit\`a di Bari$^{b}$, I-70126 Bari, Italy }
\author{G.~Eigen}
\author{B.~Stugu}
\affiliation{University of Bergen, Institute of Physics, N-5007 Bergen, Norway }
\author{D.~N.~Brown}
\author{L.~T.~Kerth}
\author{Yu.~G.~Kolomensky}
\author{G.~Lynch}
\affiliation{Lawrence Berkeley National Laboratory and University of California, Berkeley, California 94720, USA }
% ...
\collaboration{The \babar\ Collaboration}
\noaffiliation

\date{July 3rd 2012}

\begin{abstract}

The photon spectrum in the inclusive electromagnetic radiative decays of the \PB meson,  \bxsg plus \bxdg, is studied  using a data sample 
of $\lumibbpm$ $\FourS \to \BB$  decays collected by the \babar\ experiment at SLAC. The 
spectrum is used to extract the  branching fraction $\BR(\bxsg)=\PBFresultoneeightone$ for $\eg>1.8 \gev$
and the direct $\CP$ asymmetry $\acp(\bxsdg)=\acpresult$. The effects of detector resolution and Doppler 
smearing are unfolded to measure the photon energy spectrum in the \PB meson rest frame.

\end{abstract}

\pacs{13.25.Hw, 12.15.Hh, 11.30.Er}
\maketitle

In the standard model (SM), the  electromagnetic radiative decays of the \Pqb quark,  \bsg and \bdg , 
proceed via a loop diagram at leading order. A wide variety of new physics (NP) scenarios  such as supersymmetry 
may cause new contributions to the loop~\cite{Bertolini:1987pk,Hewett:1996ct,Carena:2000uj,Baer:2003yh,Huo:2003vd,
Buras:2002vd,Frank:2002nj,Agashe:2001xt} at the same order as the SM, resulting in significant deviations 
for both the branching fractions and the direct \CP asymmetry
\begin{equation*} 
  \acp = \frac{\Gamma[\bTosplusdGamma]-\Gamma[\bbarTosbarplusdbarGamma]}
       {\Gamma[\bTosplusdGamma]+\Gamma[\bbarTosbarplusdbarGamma]}.
\end{equation*}
Inclusive 
hadronic branching fractions (BF) $\BR(\bxsg)$ and $\BR(\bxdg)$ can be equated with the perturbatively calculable partonic BF $\BR(\bsg)$ and $\BR(\bdg)$  at the level of a few percent~\cite{Bigi:1992su}, allowing
theoretically clean predictions. At next-to-next-to-leading-order (four-loop), the  SM calculation 
gives  $\BR(\bxsg) = (3.15 \pm 0.23) \times 10^{-4}$\,($\eg > 1.6\gev$)~\cite{Misiak:2006zs}, 
where \eg is the photon energy measured in the rest frame of the \PB meson. $\BR(\bxdg)$ is suppressed by a factor of $|V_{td}/V_{ts}|^2 \approx 0.04$, where $V_{ij}$ are the elements of the Cabbibo-Kobayashi-Mashawa(CKM) quark-mixing matrix. NP with non-minimal flavor violation can also significantly enhance $\acp$~\cite{Hurth:2003dk}, which is approximately $ 10^{-6}$ in the SM \cite{Soares:1991te,Hurth:2001yb,Benzke:2010tq}. 
Consequently the precision measurement of these decays has long been identified as important in the search for NP. They are central to the program of the future \superb factories~\cite{Browder:2008em,O'Leary:2010af,Aushev:2010bq}, which will probe NP mass scales up to 100 TeV. 

In this letter, new precise measurements of $\BR(\bxsg)$ and $\acp$ are
presented. The analysis has been significantly improved from our previous result~\cite{Aubert:2006gg},
which it supersedes. In addition, the  shape of the photon energy spectrum is measured in the \PB meson rest frame. It is insensitive to 
NP~\cite{Kagan:1998ym} but can be used to  determine the Heavy Quark Expansion parameters  \mb and 
\mupsq ~\cite{Benson:2004sg,Neubert:2005nt},  related to the mass and momentum 
of the  \Pqb quark within the $\PB$ meson. These parameters are used to reduce the uncertainty 
in the extraction of the CKM elements $\Vcb$ and $\Vub$ from semi-leptonic  \PB meson 
decays~\cite{Bauer:2004ve,Lange:2005yw,Bauer:2002sh,Gambino:2007rp}.

This Letter summarizes a fully inclusive analysis of $\bxsg$ decays  collected from $\epem\to\FourS\to\BB$ events.
Full details are given in Ref.~\cite{bib:ourPRD}.
The photon from the decay of one \PB meson is measured, but \xs is not reconstructed.  This avoids  
large uncertainties from the modeling of the \xs system, at the cost of large backgrounds, which need 
to be strongly suppressed. The principal backgrounds are from  other \BB decays containing a  high-energy photon 
and from continuum $\Pq\Paq$ ($q=udsc$) and $\Pgtp\Pgtm$ events. The continuum background, including a 
contribution from initial-state radiation, is suppressed principally by requiring a high-momentum charged lepton (``lepton tag'') from
the non-signal \PB decay,  and also by discriminating against events with a  more jet-like topology. The \BB  background to high-energy photons, dominated by $\Pgpz$ and $\Pgh$ decays, is reduced by vetoing reconstructed $\Pgpz$ or $\Pgh$ mesons.
The residual continuum background is subtracted using off-resonance data collecdted at a center-of-mass (CM)  energy 40\mev  
below the \FourS, while the  remaining \BB background is estimated  using a Monte Carlo (MC) simulation that has 
been corrected  using data control samples. The photon energy spectrum is measured in the \FourS rest frame.
Quantities measured in this frame are denoted by an asterisk, \forex, \egcms.

The data were collected with the \babar\ detector~\cite{Aubert:2001tu} at the PEP-II
asymmetric-energy \epem collider. 
The on-resonance integrated luminosity is \onlumi, corresponding to \lumibbpm \BB events. Additionally, \offlumi  of 
off-resonance data  are used. The \babar\ MC simulation, based on 
\textsc{GEANT4}~\cite{Agostinelli:2002hh}, \textsc{EVTGEN}~\cite{Lange:2001uf} and \textsc{JETSET}~\cite{Sjostrand:1993yb}, is used to 
generate samples of \BpBm and \BzBzb (excluding signal channels), $\Pq\Paq$ , $\Pgtp\Pgtm$, and signal events. 
The signal models used to compute efficiencies are based on QCD calculations in the ``kinetic scheme'' ~\cite{Benson:2004sg}, ``shape function scheme''~\cite{Neubert:2005nt},~and in an earlier model~\cite{Kagan:1998ym}. These calculations 
approximate the \xs resonance structure with a smooth distribution in the hadronic mass  \mxs. The portion of the \mxs spectrum below 1.1\gevcc, where
the $\PKst(892)$ dominates, is replaced by a Breit-Wigner $\PKst(892)$ distribution. The analysis is performed ``blind'' in the range 
$1.8 <\egcms <2.9 \gev$; that is, the on-resonance data are
not examined until all selection requirements are finalized and  the corrected \BB backgrounds determined. The signal range 
is limited by large \BB backgrounds at low \egcms.

The event selection begins by requiring at least one photon candidate with $1.53<\egcms<3.50\gev$. 
A photon candidate is an electromagnetic calorimeter (EMC) energy cluster with a lateral profile consistent 
with that of a single photon, isolated by 25 \cm from any other cluster, and  
well contained in the calorimeter.
Photons that are consistent with originating from an identifiable \Pgpz or 
$\Pgh\to\gamma\gamma$ decay are vetoed.  Hadronic events are selected by requiring at least three  reconstructed 
charged particles and the normalized second Fox-Wolfram moment  $R_2^*$ to be less than 0.9. To reduce radiative 
Bhabha and two-photon  backgrounds, the number of charged particles plus half the number of photons with energy 
above 0.08\gev is required to be at least $ 4.5$.

About 20\% of \PB mesons decay semileptonically to either $e$ or $\mu$. Leptons from these decays are emitted 
isotropically and tend 
to have higher momentum than the continuum background in which the lepton and photon candidates also tend 
to be anti-collinear. A tagging lepton ($\ell=e,\mu$) is required to have  momentum $p^{*}_{l}>1.05\gevc$ and 
an angle relative to the photon  $\cos\theta^{*}_{\gamma\ell}> -0.7$ to suppress this background. It does not 
compromise the inclusiveness of the \bxsg\ selection since it comes from the recoiling \PB meson.  The presence 
of a relatively high-energy neutrino in semileptonic \PB decays is used to further suppress the background by 
requiring the missing energy of the event to satisfy
$E^{*}_\mathrm{miss}>0.7\gev$.

The sample is separated into electron and muon tags. For each, $p^{*}_{l}$ and $\cos\theta^{*}_{\gamma\ell}$
are then combined in a neural network (NN) with eight event-shape variables that exploit the difference in topology 
between  isotropic \BB events and jet-like continuum events. 
The NN is trained to separate
signal-like events from continuum background using MC samples. 
The \BB background sample is excluded from the training
because it is used for background subtraction and is
topologically similar to the signal.
 
The selection criteria are optimized for statistical precision. This was done 
iteratively for five variables:  the two NN outputs, the energies of the lower-energy 
photon in the \Pgpz and \Pgh vetoes, and $E^{*}_\mathrm{miss}$.
The signal efficiency for the entire selection depends on \egcms, falling  at lower values.
This effect is significantly reduced from our previous analysis, lessening the uncertainty due to the
assumed signal model (``model-dependence''). The efficiency integrated over the range $1.8 < \egcms < 2.8 \gev$  is about 2.5\%, while 
only 0.0005\% of the continuum and 0.013\% of the \BB background remain in the sample.

The remaining continuum background is estimated with off-resonance data scaled to the on-resonance luminosity and adjusted
to account for the 40\mev  CM energy difference. 
The \BB background is estimated with the \BB MC sample. It consists predominantly of photons originating 
from $\Pgpz$ or $\Pgh$  decays ($\approx 80\%$ in the signal region), electrons ($\approx 10\%$)  that are 
misreconstructed, not identified,  or  undergo hard bremsstrahlung,
$\omega$ and $\Pghpr$ decays ($\approx 4\%$), and $\Pan$'s  ($\approx 2\%$) that fake photons by 
annihilating in the EMC. Each of the significant components is corrected by comparison with data control samples.

The \Pgpz and \Pgh background simulations are compared to data using the same selection criteria as for $\bxsg$ but 
removing the \Pgpz and \Pgh vetos. For this comparison the high-energy photon 
requirement is relaxed  to $\egcms > 1.03 \gev$ to increase the size of the  
sample. The yields of \Pgpz and \Pgh  are measured in bins 
of $E^{*}_{\pizeta}$ by fitting the $\gamma\gamma$ mass distributions in on-resonance data, off-resonance data, and
\BB simulation. Correction factors to the \Pgpz and \Pgh components of the \BB simulation are derived from 
these yields.  An additional correction is applied to account for data-MC differences in the low-energy photon 
detection efficiency. This has an opposite effect on the control-sample \Pgpz and \Pgh selection than on 
the standard event selection, where finding a \Pgpz or \Pgh results in the event being vetoed. 

As an antineutron control  sample could not be isolated, this source of \BB background is corrected 
by comparing in data and simulation the  inclusive antiproton yields in \PB  decay and the  EMC response 
to $\Pap$'s, using  $\overline{\Lambda}\to \overline{p}\Pgpp$ samples. The misreconstructed electron 
background is measured using $\PB \to X \PJgy(\Pep\Pem)$ data. This sample 
closely models the particle multiplicity in  $\bxsg$ events. 
Bremsstrahlung in the detector 
is reliably simulated by \textsc{GEANT4}, so no correction is necessary. 
The small contributions 
from $\omega$ and $\Pghpr$ decays are corrected in bins of \egcms using  inclusive \PB decay data. 
Nearly all of the tagging leptons arise from  $B\to
X_c\ell\nu$. The yield of such events in the  simulation is corrected as a function of lepton 
momentum according to previous \babar\ measurements~\cite{Aubert:2004td,Aubert:2009qda}.
The complete \BB background estimation
incorporates  the correction factors and uncertainties and includes correlations 
between \egcms bins. The dominant uncertainties originate from the $\Pgpz, \Pgh$, and
misreconstructed electron corrections.

\begin{figure}[htb]
\begin{center}   
  \includegraphics[width=.5\textwidth]{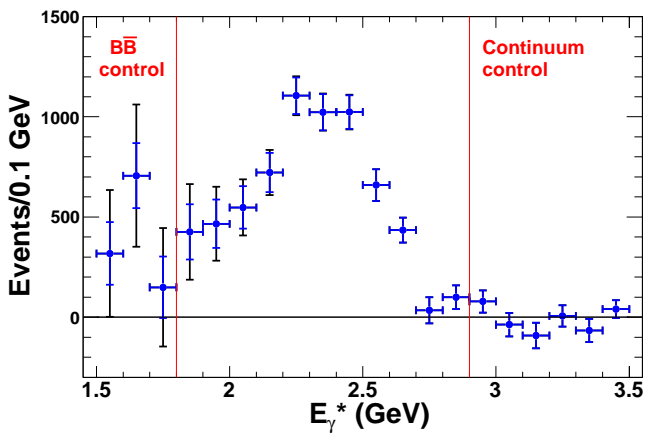}   
\end{center}
\vspace{-0.2in}
\caption{The measured \egcms photon energy spectrum after background subtraction, uncorrected for efficiency and resolution smearing. The 
inner error bars are  statistical only, while the outer include systematic errors added in quadrature. }
\label{fig:spectrumuncorrected} 
\end{figure}

Figure~\ref{fig:spectrumuncorrected}  shows the measured \egcms spectrum after subtracting both continuum and
\BB backgrounds. The systematic errors are due to the \BB subtraction uncertainty. The region 
$1.53 < \egcms < 1.80 \gev$ is dominated by \BB background, while 
the higher-energy range $2.9 < \egcms < 3.5\gev $ contains only continuum background. These regions
are used to validate the background subtraction procedure. 
In the higher-energy range there are  $ -100 \pm 138 (\mathrm{stat})$
events. In the lower-energy region there are $ 1252
\pm 272 (\mathrm{stat}) \pm 841 (\mathrm{syst})$ events. Allowing for an average of 275  signal events from a 
range of plausible signal models, and for correlations between the bins, the latter result
is consistent with zero to within one standard deviation ($1\sigma$).

To extract BF's and the shape of the spectrum, it is necessary to first correct 
for efficiency. Theoretical predictions are made for the true \eg in the \PB meson rest frame, whereas the 
\egcms is measured in  the \FourS frame. Hence it is also necessary 
to  correct for the asymmetric EMC resolution and the Doppler smearing due to the 
motion of the \PB meson in the \FourS rest frame. The  efficiency and smearing corrections depend upon the 
assumed signal shape. In both the kinetic and shape function schemes this shape is  parameterized by \mb 
and \mupsq. The Heavy Flavor Averaging Group (HFAG)~\cite{TheHeavyFlavorAveragingGroup:2010qj} has extracted 
 values and uncertainties in the kinetic scheme by fitting moments of inclusive distributions in
\bxclnu decays and previous \bxsg
measurements, and has also translated them to the shape function scheme.  These results define the 
nominal signal model (kinetic scheme) used for the BF measurement, along with a model-dependence uncertainty (kinetic and shape function schemes). To provide an independent measurement of the shape of the spectrum, 
the measured spectrum is unfolded using an iterative technique that reduces sensitivity to the signal model. In this case the initial signal model and 
model-dependence uncertainty are based on the data rather than the
HFAG parameters. The effects of efficiency and smearing cancel in the \acp measurement so 
it is extracted directly from the measured \egcms yield separated  by lepton tag charge.

The BF is computed from 
\begin{equation*}
\mathcal{B}(\bxsdg) = \alpha\sig/(2N_{\BB}\sigeff),
\end{equation*}
 where
\sig is the signal yield integrated over the \egcms ranges $1.8,1.9,2.0$ to $2.8\gev$, \sigeff is
the signal efficiency, and $N_{\BB}$ is the number of \BB pairs in the sample. The factor $\alpha$, which is close 
to unity,  corrects for 
resolution and Doppler smearing and is computed  with the nominal signal model. 
The model-dependence errors on the BF associated with the efficiency and the smearing correction are fully
correlated. The results for the three energy ranges are given 
in Table~\ref{tab:results}. 
\begin{table*}[hbt]
 \label{tab:results}
 \addtolength{\extrarowheight}{1.7pt}
\hspace*{1.0cm}
 \begin{center}
  \caption{The measured BF, first, and second moments 
  ($\pm \mathrm{stat} \pm \mathrm{syst} \pm \mathrm{model}$) for different ranges of $\eg$ in the 
  \PB rest frame. Correlations between the energy ranges are given in Ref.~\cite{bib:ourPRD}.}
  \begin{tabular*}{17.8cm}{@{\extracolsep{\fill}}ccccc} \hline\hline
   \quad \eg Range~(\gev) & $\BR(\bxsg)~(10^{-4})$                &   $\efmbrest~(\mathrm{GeV})$             &  $\varbrest ~(\mathrm{GeV}^{2})$        \\ \hline
   \quad 1.8 to 2.8   & $3.21 \pm 0.15 \pm 0.29 \pm 0.08$    & $2.267\pm 0.019 \pm 0.032 \pm 0.003$    & $0.0484 \pm 0.0053 \pm 0.0077 \pm 0.0005$   \\       
   \quad 1.9 to 2.8   & $3.00 \pm 0.14 \pm 0.19 \pm 0.06$    & $2.304\pm 0.014 \pm 0.017 \pm 0.004$    & $0.0362 \pm 0.0033 \pm 0.0033 \pm 0.0005$   \\       
   \quad 2.0 to 2.8   & $2.80 \pm 0.12 \pm 0.14 \pm 0.04$    & $2.342\pm 0.010 \pm 0.008 \pm 0.005$    & $0.0251 \pm 0.0021 \pm 0.0013 \pm 0.0009$  \\ \hline  \hline    
  \end{tabular*}
 \end{center}
 \vspace{-0.1in}
\end{table*}
The BF's have been corrected 
by a factor $1/(1 + (\Vtd/\Vts)^2) = 0.958 \pm 0.003$~\cite{Nakamura:2010zzi} to remove the contribution from $\bdg$.
The most significant systematic error is from the corrections to the $\BB$ background simulation, which in the range  $1.8\gev < \eg < 2.8 \gev$  contributes 7.8\% to a total systematic uncertainty of 9.0\%. Additional contributions added in quadrature, 
all energy-independent, arise from uncertainties in the selection efficiency (3.1\%), predominantly due to the high-energy photon and NN selections, the  semileptonic BF for \PB meson decays, and the modeling of the $X_{s}$ system. Correlations between the \BB and the signal efficiency
systematic errors contribute  an additional 2.9\% uncertainty. Finally, there is a 1.1\% uncertainty in $N_{\BB}$.

\begin{figure}[htb]
\begin{center}   
  \includegraphics[width=.5\textwidth]{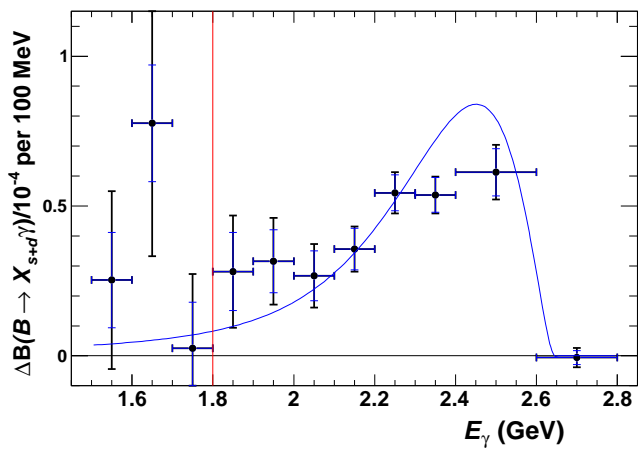}   
\end{center}
\vspace{-0.2in}
\caption{The \eg photon energy spectrum corrected for efficiency, resolution, and Doppler smearing, shown as a partial
branching fraction $\Delta \BR$. The inner error bars 
are  statistical and the outer include systematic errors added in quadrature. The vertical line shows the
boundary between the lower control region and the signal region. The curve
is the kinetic scheme model using HFAG world average parameters, normalized to data in the range $1.8 <\egb < 2.8\gev$.}
\label{fig:spectrumcorrected} 
\end{figure}

To obtain an \eg spectrum in the \PB rest frame, the \egcms spectrum shown in Fig.~\ref{fig:spectrumuncorrected} 
is corrected for selection efficiency, and the resolution
smearing and Doppler smearing are  unfolded. A simplified version~\cite{bib:2009fg} of an iterative unfolding technique~\cite{Malaescu:2009dm} is used. The method starts with an initial signal model that, when passed through the detector simulation
and event selection, closely resembles the data (shape function scheme with $\mb=4.51\gev ,\mupsq =0.46 \gev^{2}$). 
This model is used to correct for efficiency and unfold the data. A fraction, determined by a bin-dependent regularization function, of the difference between the unfolded data and
the initial signal model is used to adjust the signal model, and the process is iterated until it converges.
Only one iteration is necessary. The results are shown in Fig.~\ref{fig:spectrumcorrected}.
This technique preserves fluctuations in the spectrum and reduces the 
model error. The model dependence uncertainty is
computed using an initial model that is approximately $1\sigma$
lower than the data in Fig.~\ref{fig:spectrumuncorrected} in the region with significant \BB background 
($1.8<\egcms<2.1\gev$). The error is the absolute value of the difference bin by bin after unfolding. It
is small except near the kinematic limit,
$\eg \approx m_{\PB}/2$, where the sharply falling edge leads to strongly anti-correlated differences in
adjacent bins. To reduce this effect, the 100-MeV bins between 2.4 and 2.8\gev are combined into
200-MeV bins.  The spectral shape and the full covariance matrix, provided in Ref.~\cite{bib:ourPRD}, are used  to compute the first and second moments in Table~\ref{tab:results}. They can also be used to fit any theoretical prediction for the spectral shape. The BF's computed from the
sum of the $\Delta \BR$ in Fig.~\ref{fig:spectrumcorrected} are consistent with the values given in Table~\ref{tab:results}~\cite{bib:ourPRD}.

Finally the \egcms sample is divided into $\PB$ and $\PaB$ decays, using the charge of the lepton tag, to measure 
$\amcp(\PB \to X_{s+d}\gamma) = (N^{+}-N^{-})/(N^{+}+N^{-})$  where $N^{+(-)}$ are the 
positively (negatively) tagged signal yields. $\acp$ is then given by $\acp=\amcp/(1-2\omega)$ where $\omega$ is 
the mistag fraction. To maximize the statistical precision a requirement of  $2.1<\egcms<2.8\gev$ is made. 
This is determined from simulation and does not bias the SM prediction for the asymmetry~\cite{Kagan:1998bh}. The yields are $N^{+}=2620 \pm 158 (\mathrm{stat})$ and $N^{-}=2389 \pm 151(\mathrm{stat})$. 
The bias on \acp due 
to  charge asymmetry in the detector response or \BB background is measured to be $\Delta\amcp(\bxsdg)=-0.004 \pm 0.013$, 
using events in the \BB control region to check for a background asymmetry, and using several event samples 
($\epem\to\epem\Pgg$, $\epem\to\mmg$ and $\PB \to K^{(*)}\PJgy(\ell^{+}\ell^{-})$)
to check for a lepton tag asymmetry. The mistag fraction  $\omega = 0.133\pm0.006$ is dominated by  $\PBz\PaBz$ mixing, which contributes  $0.093 \pm 0.001$~\cite{Nakamura:2010zzi}, with an additional $0.040 \pm 0.005$ arising from wrong-sign leptons from the \PB decay chain and from misidentifcation of hadrons as leptons.  After correcting for charge bias and mistagging it is found
\begin{equation*}
  \acp  = 0.057 \pm 0.060 (\mathrm{stat}) \pm 0.018 (\mathrm{syst}) .
\end{equation*}
The systematic error includes relative
uncertainties from the \BB background subtraction (2.2\%) and mistagging (1.8\%). The uncertainty due to differences in the \bxsg and \bxdg spectra is  negligible. 

In summary, the photon spectrum of $\bxsdg$ decays has been measured and used to extract the
branching fraction, spectral moments, and \acp.
Previous inclusive measurements of $\bxsg$ have been presented by the CLEO~\cite{Chen:2001fja}, \babar\ ~\cite{Aubert:2006gg}, and Belle~\cite{Limosani:2009qg} Collaborations.
The measured branching fraction $\BR(\bxsg)=(3.21 \pm 0.15 \pm 0.29 \pm 0.08)\times 10^{-4}$ $(1.8<\eg<2.8 \gev)$ 
is comparable in precision to the Belle result,
 $(3.36 \pm 0.13 \pm 0.25 \pm 0.01)\times 10^{-4}$, but with a dataset that has $60 \%$ smaller integrated luminosity. 
The BF for $1.8 < \eg < 2.8 \gev$ is extrapolated to the range $\eg > 1.6 \gev$ using a factor of $1/(0.968\pm 0.006)$
 determined by HFAG. This results in  $\BR(\bxsg) = (3.31 \pm 0.16 \pm 0.30 \pm 0.09) \times 10^{-4}$ for  $\eg > 1.6\gev$, in good agreement with the SM prediction. 
The extrapolated $\BR(\bxsg)$ can be used to constrain NP. For 
example, in a type-II two-Higgs-doublet model~\cite{Misiak:2006zs,Haisch:2008ar} the 
region $M_{H^{\pm}}<327 \gev$ is excluded independent  of $\tan{\beta}$ at 95\% confidence level. This limit is far more 
stringent than that from direct searches at the LHC~\cite{Chatrchyan:2012cw,Atlas:2012q}. The \acp measurement 
is the most precise to date and can be used to constrain non-minimal flavor-violating models~\cite{Hurth:2003dk}. The
measured moments and spectra provide input to improve the precision on the HFAG estimation of  $\mb$ and $\mupsq$, which will
result in a reduced error on $|V_{ub}|$. Finally, the improved technique presented in this paper can 
be applied with increased precision at future \superb factories.

 We are grateful for the excellent luminosity and machine conditions
provided by our \pep2\ colleagues, 
and for the substantial dedicated effort from
the computing organizations that support \babar.
The collaborating institutions wish to thank 
SLAC for its support and kind hospitality. 
This work is supported by
DOE
and NSF (USA),
NSERC (Canada),
CEA and
CNRS-IN2P3
(France),
BMBF and DFG
(Germany),
INFN (Italy),
FOM (The Netherlands),
NFR (Norway),
MES (Russia),
MICIIN (Spain),
STFC (United Kingdom). 
Individuals have received support from the
Marie Curie EIF (European Union)
and the A.~P.~Sloan Foundation (USA).

 \bibliography{note2429}
 \bibliographystyle{apsrev}

\end{document}